\documentclass[aps,prb,twocolumn,superscriptaddress,showpacs]{revtex4-1}

\usepackage{graphicx}
\usepackage{dcolumn}
\usepackage{multirow}
\usepackage{amsmath,bm}
\usepackage{color}%

\begin{document}

\title{Is hexagonal boron nitride always good as a substrate for 
carbon nanotube-based devices?}

\author{Seoung-Hun Kang}
\affiliation{Department of Physics and
             Research Institute for Basic Sciences,
             Kyung Hee University, Seoul, 130-701, Korea}

\author{Gunn Kim}
\email[Corresponding author. E-mail: ]{gunnkim@sejong.ac.kr}
\affiliation{Department of Physics, 
             Graphene Research Institute and 
             Institute of Fundamental Physics,
             Sejong University, Seoul, 143-747, Korea}

\author{Young-Kyun Kwon}
\email[Corresponding author. E-mail: ]{ykkwon@khu.ac.kr}
\affiliation{Department of Physics and
             Research Institute for Basic Sciences,
             Kyung Hee University, Seoul, 130-701, Korea}

\date{\today}

\begin{abstract}
Hexagonal boron nitride sheets have been noted especially for their
enhanced properties as substrates for $sp^2$ carbon-based nanodevices.
To evaluate whether such enhanced properties would remain under
various realistic conditions, we investigate the structural and
electronic properties of semiconducting carbon nanotubes on perfect
and defective hexagonal boron nitride sheets under an external
electric field as well as with a metal impurity, using density
functional theory. We verify that the use of a perfect hexagonal boron
nitride sheet as a substrate indeed improves the device performances
of carbon nanotubes, compared with the use of conventional substrates
such as SiO$_2$. We further show that the hexagonal boron nitride even
with some defects can perform better performance as a substrate. Our
calculations, on the other hand, also suggest that some defective
boron nitride layers with a monovacancy and a nickel impurity could
bring about poor device behaviors since the imperfections impair
electrical conductivity due to residual scattering under an applied
electric field.
\end{abstract}


\maketitle

\section{Introduction}
\label{Introduction}

The search for substrates to improve the performance of nanoelectronic
devices has been an important research topic. Thus far,
metals,~\cite{{GRARU},{GRAIR}} mica,~\cite{GRAMICA}, SiC\cite{{SIC1},
{SIC2}} or SiO$_2$~\cite{{SIO21},{SIO22},{SIO23},{SIO24},{SIO25}}
have been used as substrates. For carbon-based devices, SiO$_2$ has
been most commonly used as a substrate. Although the use of SiO$_2$
provides many advantages, its primary drawback is to reduce the
electronic mobility due to charge density fluctuations induced by the
impurities presented in devices.~\cite{{REDM},{REDM1}}

Recently, hexagonal boron nitrides (hBN) began to take center stage
as a better substrate for graphene-based nanodevices than conventional
substrates, such as SiO$_2$. It has been demonstrated that graphene on
the hBN substrate exhibits increased mobility,~\cite{{IMMO},{HQGRA}}
significant improvements in quantum Hall measurements,~\cite{HQGRA}
and enhancement of graphene nanodevice reliability.~\cite{exper1} It
has also been shown that carbon nanotubes (CNTs) based devices on hBN
substrates exhibit better device characteristics than on conventional
substrates,~\cite{APL2014} similarly as graphene-based devices.
According to studies based on scanning tunneling microscopy
measurements, hBN provides an extremely flat surface with
significantly less electron-hole puddles than SiO$_2$.~\cite{{STM1},
{STM2}} In addition to the advantages of providing an atomically smooth
surface (relatively free of dangling bonds and charge traps), a large
band gap of ${\agt}5$~eV,~\cite{BG} chemical inertness, and a low
density of charged impurities, hBN sheets also exhibit an unusual
electronic structure that is not observed in most wide band gap
materials.

It used to be difficult to fabricate perfect hBN sheets, particularly
with a large area, but recently, large-area hBN sheets have been grown
by using chemical vapor deposition (CVD).~\cite{{BF3NH3CVD},
{BClNH3CVD},{B2H6NH3CVD},{B3N3H6CVD},{B3N3H3Cl3CVD1},{B3N3H3Cl3CVD2},
{B3N3Cl3CVD},{FLBNCVD},{MonoBNCVD}} A recent intriguing experiment
showed that a direct CVD growth of single-layer graphene on a
CVD-grown hBN film exhibits better electronic properties than that of
graphene transferred on the hBN film.~\cite{MWang} Thanks to this
method, one need not worry about the polymer residues remaining on
transferred graphene any longer. However, there still remain two
concerns that CVD-grown hBN sheets may possess intrinsic defects, such
as vacancies,~\cite{CJin,VACCVD} and carbon nanotubes (CNTs) may
contain metal impurities used as catalysts during their
growth,~\cite{{IM1},{IM2},{IM3},{IM4},{IM5}} which were further
explored to confirm the usefulness of hBN substrates.

In this paper, we report a first-principles study of the structural
and electronic properties of a semiconducting CNT on defective hBN
sheets in the presence of an external electric field (E-field), as
well as on a perfect hBN sheet for comparison. For a CNT on a perfect
hBN, the E-field shifts the electronic states from the hBN relatively
to those from the CNT or to the Fermi level, but the hBN states are
still far from $E_F$, and the CNT states are not altered by the
E-field considered. Thereby the perfect hBN may lead to improvement in
device performance, compared with conventional substrate materials
such as SiO$_2$, as confirmed by experiments.~\cite{{IMMO},{HQGRA},
{HQGRA}, {STM1},{STM2},{MWang},{exper1}} For a CNT on a defective hBN,
on the other hand, our study shows that the electronic states
originating from the hBN substrate are shifted to near $E_F$ due to a
nickel impurity, which was selected as an exemplary catalyst for CNT
growth, and a vacancy in the hBN sheet. Such shifted states and the
in-gap states from defects could result in electronic scattering near
the Fermi level ($E_F$) or unwanted electrical conduction (leakage
current) through the hBN substrate, causing some critical problems in
the CNT devices.

\section{Computational details}
\label{Computational}

We carried out first-principles calculations using the Vienna
{\em{ab initio}} simulation package (VASP).~\cite{VASP} Projector
augmented wave potentials were employed to describe the valence
electrons.~\cite{PAW} The exchange-correlation functional was treated
within the spin-polarized local density approximation (LDA) in the
form of Ceperley-Alder parametrization.~\cite{LDACA} The cutoff
energy for the plane wave basis was chosen to be 450 eV, and the
atomic relaxation was continued until the Hellmann-Feynman force
acting on every atom became lower than 0.03~eV/{\AA}. For more precise
calculations, we included the dipole correction.

We first found the equilibrium structure of the hBN sheet with the
primitive unit cell, where the B-N bond length $d_{\rm{BN}}$ was
calculated to be 1.44~{\AA} corresponding to the lattice constant of
2.50~{\AA}, which is in excellent agreement with the experimental
values of $2.49\sim2.52$~{\AA}.~\cite{CJin,Paszkowicz2002} To arrange
a $(10,0)$ CNT on the pristine hBN sheet, we prepared an orthorhombic
supercell with three side lengths of $a=4.32$~{\AA} ($=3d_{\rm{BN}}$),
$b=17.5$~{\AA}, and $c=25.0$~{\AA}, including the rectangular hBN with
two sides of $a$ and $b$ and the zigzag CNT placed on the hBN sheet
along the $a$ direction. To compensate the discrepancy of their
``native'' lattice constants along the CNT axial direction, the CNT
was elongated by ${\sim}5$~\%. The other side lengths, $b$ and $c$,
were selected large enough to avoid the intertube interaction from the
neighboring cells, and to contain a vacuum region of
${\sim}14.4$~{\AA} between the top of the CNT and the bottom of the
hBN located in the next cell above. For the systems with various
vacancies on hBN and/or a Ni impurity, we increased $a$ by a factor of
four to be $17.28$~{\AA} to ignore the interaction from those defects
located in neighboring cells. The Brillouin zone was sampled using a
$\Gamma$-centered $10{\times}1{\times}1$ ($5{\times}1{\times}1$)
$k$-point mesh for the system with a smaller (larger) $a$ value. The
electronic levels were convoluted using Gaussian broadening with all
width of 0.05~eV to obtain the DOS.

\section{Results and discussion}
\label{Results}

\begin{figure}[t]
\includegraphics[width=1.0\columnwidth]{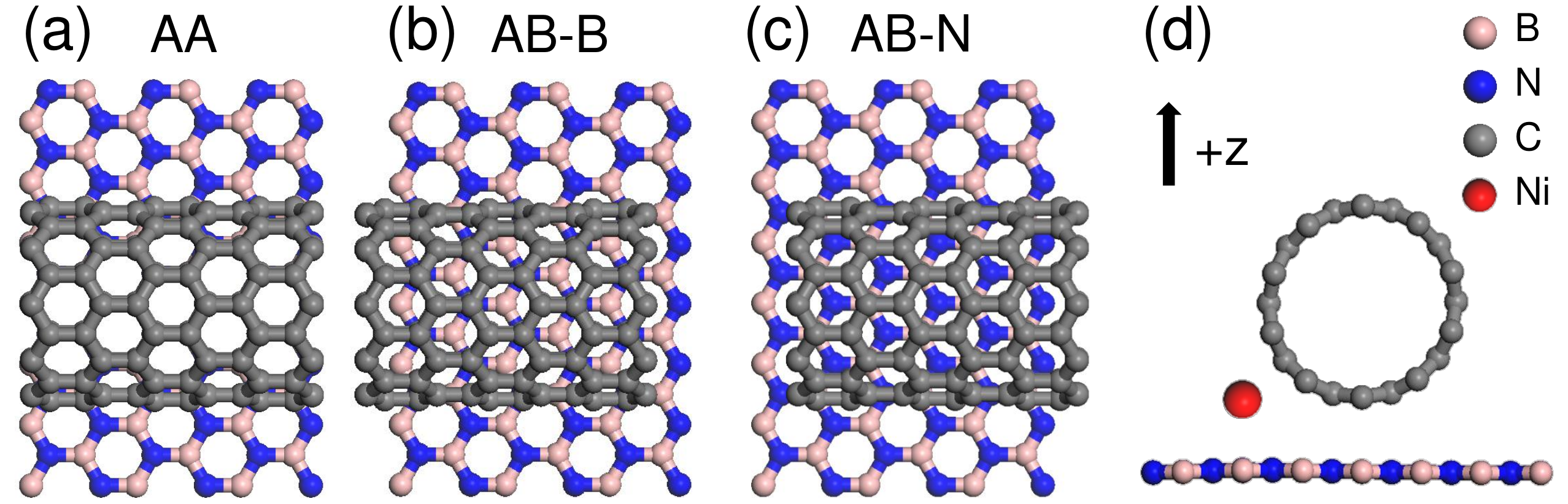}
\caption{(Color online) 
Three highly-symmetric stacking configurations of $(10,0)$ CNT
adsorbed on hBN sheet classified similarly in graphite with AA and AB
stacking configurations: (a) AA, (b) AB-B (B atoms located at the
hollow site of the CNT), (c) AB-N (N atoms at the hollow site)
stackings in top view, and (d) in side view. In (d), the arrow
indicates the $+z$ direction along which or the opposite of which an
external E-field is applied. A Ni impurity atom is also shown in (d).
The pink, blue, gray and red spheres represent boron, nitrogen, carbon
and nickel atoms, respectively.
\label{Fig1}}
\end{figure}

First, we found the equilibrium stacking configuration of a $(10,0)$
CNT adsorbed on a perfect hBN substrate. Fig.~\ref{Fig1}(a--c) show
three highly-symmetric stacking configurations of the CNT adsorbed
onto the hBN sheet similarly in graphite stacking configurations. In
AA, some of C$_6$ hexagonal rings in CNT are placed to match exactly
on the top of B$_3$N$_3$ hexagonal rings in hBN. On the other hand,
there are two possibilities in AB configuration dissimilar from AB
graphite: in AB-B, the hollow sites of the CNT are located at the top
of B atoms, while they are at the top of N atoms in AB-N. Our
calculations reveal that the most energetically favorable
configuration is AB-N stacking, which was also observed as the most
stable stacking configuration for graphene on
hBN.~\cite{Giovannetti2007} With respect to the most stable AB-N
stacking configuration, AB-B and AA have 61 and 73~meV higher
energies, respectively. Therefore, we considered only AB-N stacking
configuration for further investigations. Its projected density of
states (PDOS) were calculated to study the effects of an external
E-field and a Ni impurity on the electronic structure. The latter was
considered since nanoparticulate nickel is often used as a catalyst to
synthesize CNTs, and thus the system may contain nickel impurity
atoms,~\cite{NiCNT} unless Ni nanoparticles have been completely
removed from the CNT surface or substrate after the growth.

\begin{figure}[t]
\includegraphics[width=1.0\columnwidth]{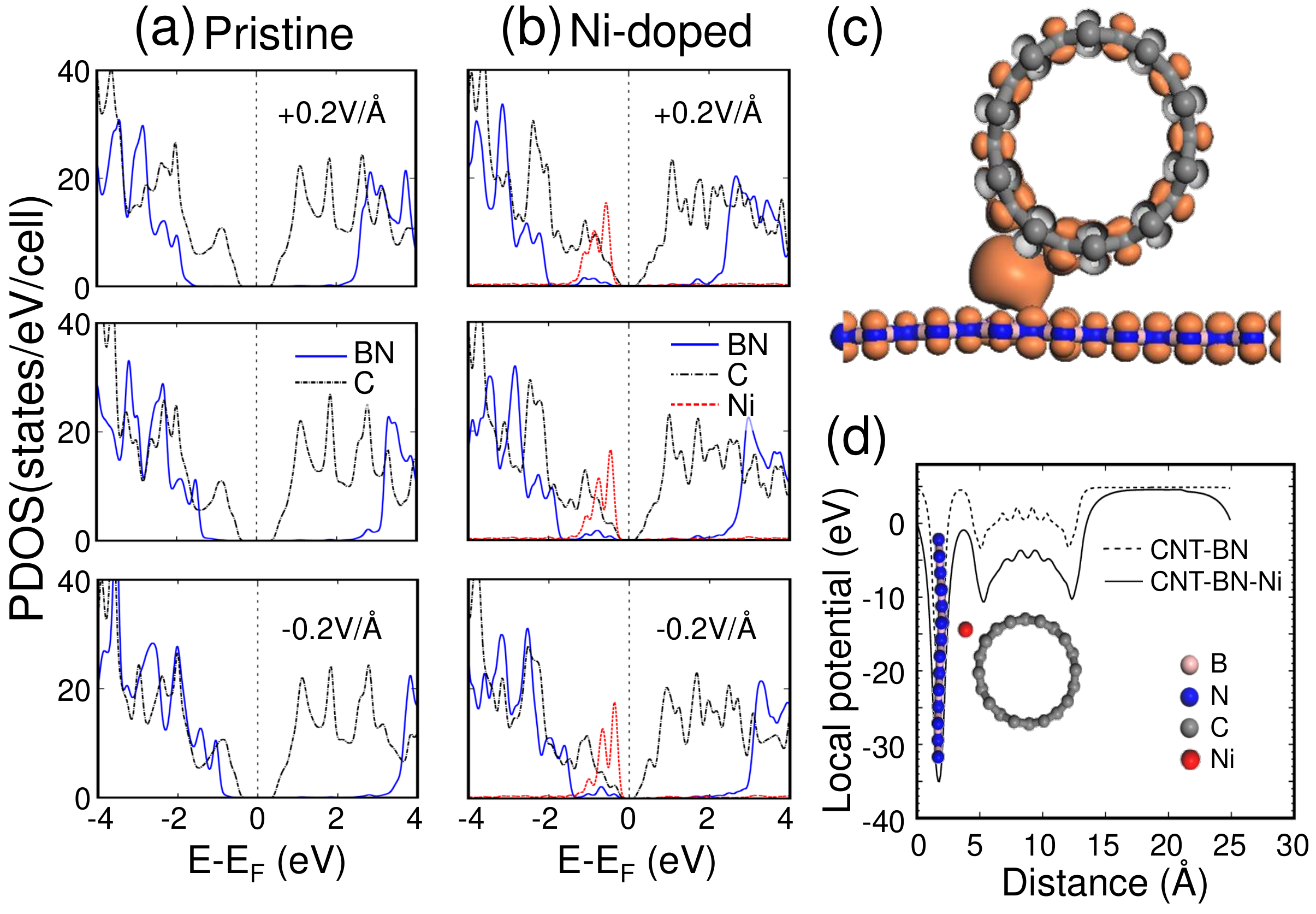}
\caption{(Color online) 
PDOSs of $(10,0)$ CNTs deposited on (a) pristine and (b) Ni-doped hBN
sheets in the absent (middle) and the presence of an E-field applied
with 0.2~V/{\AA} along the $+z$ (top) and $-z$ (bottom) directions,
which are normal to the hBN sheet. In each graph in (a) and (b), a
blue solid, a black dash-dotted, and a red dashed lines represent the
PDOSs of the hBN, the CNT, and the Ni atom, respectively. The Gaussian
broadening of 0.15~eV was used for all PDOSs. There is an
arbitrariness to set the Fermi level ($E_F$) anywhere within the
energy gap, we set $E_F$ at the center of the energy gap.(c) Local
charge density plotted within the energy window where the strong Ni
peaks is observed in the middle graph of (b). (d) Local potential (LP)
averaged over a plane parallel to hBN sheet as a function of the
distance perpendicular to the sheet. A solid (dashed) line is for
the system with (without) the Ni impurity.
\label{Fig2}}
\end{figure}

Fig.~\ref{Fig2}(a) shows the PDOS of the pristine AB-N configuration
in the absence and the presence of an external E-field. As observed in
the middle graph, the conduction band minimum (CBM) and the valence
band maximum (VBM) of the hBN, which are respectively located
${\sim}2.7$~eV above and ${\sim}1.5$~eV below $E_F$ in the absence of
the E-field, are quite far from the CBM and the VBM of the CNT,
respectively. As shown in the top (bottom) graph, the external E-field
applied along the $+z$ ($-z$) direction shifts the hBN energy bands
down (up) noticeably toward the lower (higher) energy relative to
those of the CNT due to lower (higher) electrostatic potential in the
hBN side. Up to the field strength of 0.2~V/{\AA}, however, the CBM
and VBM of the hBN are located at least 2~eV above and 1~eV below
$E_F$, respectively. Therefore, we conclude that none of the hBN
states affect the electronic conduction, regardless of the external
E-field, and thus the conduction may occur only through the CNT.

Our results confirm the experimental observations~\cite{{IMMO},
{HQGRA},{HQGRA},{STM1},{STM2}} of hBN sheets being much better
substrates than conventional ones, such as SiO$_2$, for CNT or
graphene~\cite{Park2014a} based single-gated field effect
transistors (FETs). In a single-gated FET, an applied gate bias
generates an E-field between the gate and the channel moving either
the VBM or CBM of the hBN close to the counterpart of the CNT, but the
energy spacing between them is still kept large, as shown in
Fig.~\ref{Fig2}(a). In a dual-gated FET, however, one gate can be used
to control the relative position of the VBM or the CBM of the hBN,
while the other to adjust the chemical potential of the CNT channel.
This could cause an unexpected output of conduction through not only
the CNT channel, but also the hBN substrate.

Fig.~\ref{Fig2}(b) shows the PDOS of the Ni-doped AB-N configuration
in the absence and the presence of an E-field. Similar to the pristine
case, the applied E-field shifts the hBN states up or down by the
field directions while remaining them still far from $E_F$. Just below
$E_F$, however, are there some strongly localized peaks originating
from the Ni $3d$ orbitals, which show no practical change under the
applied E-field. In addition, we observed small ``satellite'' states
from the hBN induced by the Ni states and a little modification in the
CNT states just below $E_F$. Our Bader charge analysis showed that
$0.26~e$ and $0.07~e$ have been transferred from the Ni adatom to the
CNT and to the hBN, respectively. We also calculated the local charge
density corresponding to the localized peaks from the Ni $3d$. As
displayed in Fig.~\ref{Fig2}(c), it exhibits a charge overlap between
the CNT and the hBN through the Ni impurity indicating that the Ni
adatom mediates a coupling between the CNT and the hBN sheet. This
implies that the nickel atom may play a crucial role as a scattering
center when the Fermi level is shifted into this energy window via
other doping or by applying a gate E-field.

Local potential (LP) was also calculated for the Ni-doped CNT-on-hBN
system, as well as the undoped system. Fig.~\ref{Fig2}(d) shows the
calculated LP averaged over a plane parallel to hBN sheet as a
function of the distance normal to the slab. We found that the LP
values of the undoped system in the region between the CNT and the hBN
appear to be almost the same as those in the vacuum region as
represented by a dashed line, indicating that the interaction between
the CNT and the hBN sheet may not influence the LP values in between.
We can, therefore, conclude that the presence of hBN sheet does not
alter the fundamental response of CNT when no metal impurity is
present. For the Ni-doped system, in contrast, its LP values, plotted
with a solid line, appear to be ${\sim}5.5$~eV lower, compared with
those of the Ni-free system with respect to their vacuum values.

Next, we explored the effects of vacancies existing in the hBN sheet.
It was reported that vacancies can be formed during the growth of hBN
sheets by the CVD methods.~\cite{VACCVD} We considered two types of
monovacancies: a boron vacancy denoted as $V_{\rm{B}}$ and nitrogen
vacancy as $V_{\rm{N}}$; and a kind of multivacancy composed of three
boron and one nitrogen empty sites ($V_{\rm{B_3N}}$), which is one of
the two smallest triangular multivacancies.~\cite{CJin} The other
triangular multivacancy consisting of one missing B atom and three
missing N atoms ($V_{\rm{BN_3}}$) was not taken into account because
experimental observations showed that more boron atoms are missing
than nitrogen atoms and most of the edge-terminating atoms around the
vacancies are doubly-coordinated nitrogen atoms.~\cite{CJin} It was
found that the CNT prefers to be placed just above $V_{\rm{B}}$ and
$V_{\rm{N}}$ rather than on the perfect hBN region with the energy
differences of $0.10$~eV and $0.13$~eV, respectively. On the other
hand, the CNT does not prefer $V_{\rm{B_3N}}$ to a defect-free hBN
region, since the $V_{\rm{B_3N}}$ defect does not possess broken bonds
after the edge reconstruction. To investigate the vacancy effects on
the electronic structures, we only considered the configurations in
which the CNT is placed on the defect site.

\begin{figure}[t]
\includegraphics[width=1.0\columnwidth]{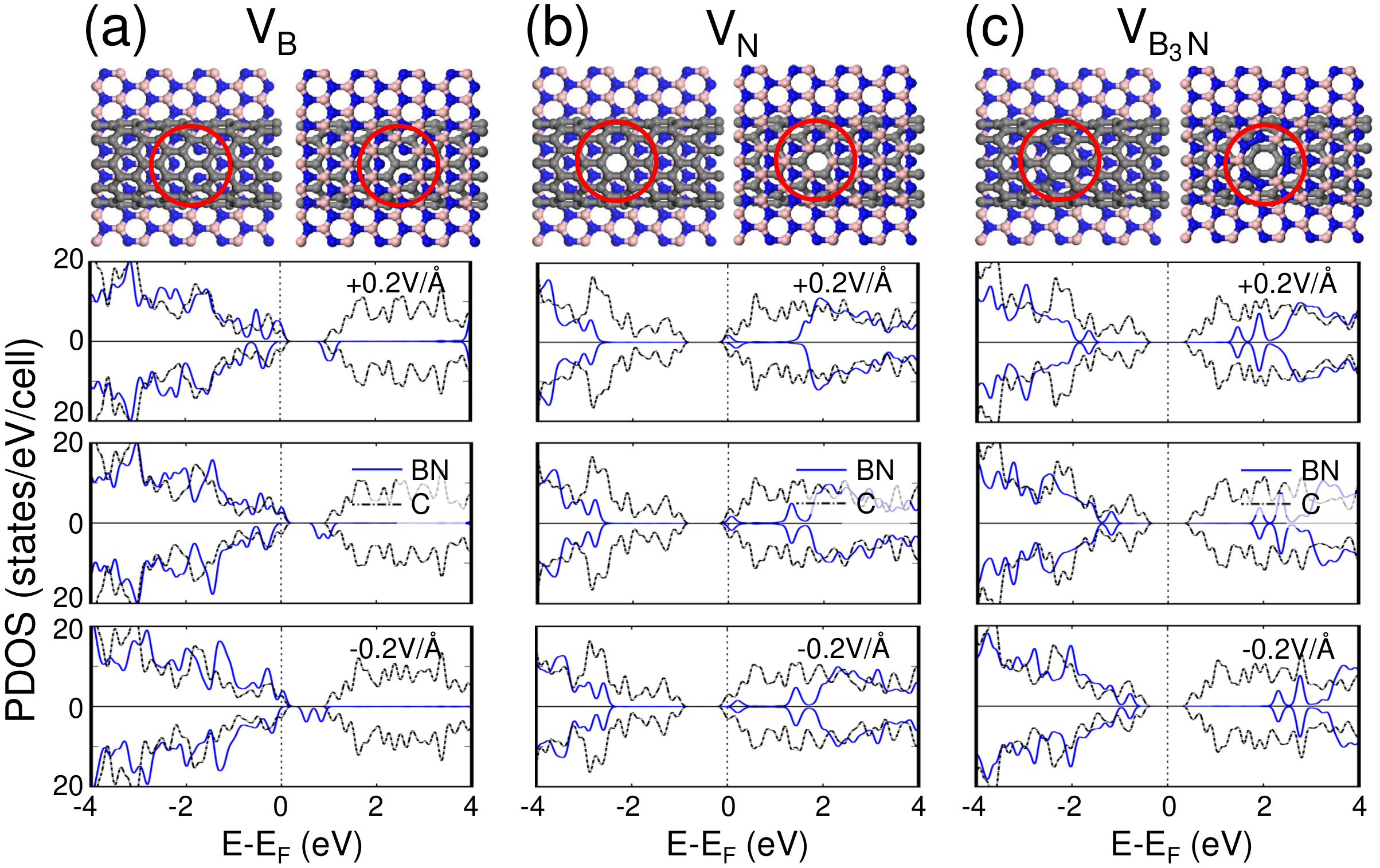}
\caption{(Color online) 
The optimized structures of $(10,0)$ CNTs placed on top of (a) a boron
monovacancy ($V_{\rm{B}}$), (b) a nitrogen monovacancy ($V_{\rm{N}}$),
and (c) a triangular multivacancy formed by three missing B and one
missing N atoms ($V_{\rm{B_3N}}$) created on hBN sheets. For each
case, the left (right) configuration is depicted in top (bottom) view.
The vacancy in each image is enclosed by the red circle. C, B, and N
atoms are denoted by the same color scheme used in Fig.~\ref{Fig1}.
Below these structures are shown their corresponding spin-resolved
PDOSs in the absence (middle) and the presence of an E-field applied
with 0.2~V/{\AA} along the $+z$ (top) and $-z$ (bottom) directions.
The Gaussian broadening of 0.15~eV was used. While the system with
$V_{\rm{B}}$ exhibits a magnetic characteristic clearly, those with
$V_{\rm{N}}$ and $V_{\rm{B_3N}}$ do not. Note in Fig.~\ref{Fig3}(b)
that an extremely small difference is recognized between majority and
minority PDOSs near $E_F$ at $|{\bf{E}}|=+0.2$~V/\AA.
\label{Fig3}}
\end{figure}



Fig.~\ref{Fig3} shows the optimized structures in top and bottom views
of the CNTs on the hBN sheet with $V_{\rm{B}}$, $V_{\rm{N}}$, and
$V_{\rm{B_3N}}$ defects. Their corresponding spin-resolved PDOSs are
also shown in the absence and the presence of an applied E-field.
Similar to the perfect hBN cases, the applied E-field shifts the hBN
states up or down relative to CNT states, depending on the field
directions. In Fig.~\ref{Fig3}(a) for $V_{\rm{B}}$, we found that the
Fermi level, $E_F$, is located just below the VBM of the CNT, which
overlaps with that of the hBN. It means that the CNT becomes weakly
hole-doped by the presence of $V_{\rm{B}}$. In addition, there exist
two localized empty states originating from $V_{\rm{B}}$ just above
$E_F$, which exhibit two interesting behaviors. One is that, in
response to the E-field, they move in the opposite direction of the
other hBN states, i.e., up (down) in energy with the positive
(negative) E-field along the $z$-direction. The other is that these
localized states, which are almost completely degenerate at an E-field
value of $+0.2$~V/{\AA}, become split into two separate states at a
value of $-0.2$~V/{\AA}. These interesting behaviors are accounted for
as following. Were it not for the CNT, $V_{\rm{B}}$ would possess its
three-fold symmetry with its three equivalent unsaturated N atoms. In
the presence of the CNT, however, the interaction with the CNT, which
is very weak, but not negligible, provides a small perturbation
breaking its three-fold symmetry lifting its degeneracy. As displayed
in PDOSs of Fig.~\ref{Fig3}(a), such perturbation becomes weaker and
stronger at E-field values of $+0.2$~V/{\AA} (top) and $-0.2$~V/{\AA}
(bottom), respectively, than at no E-field (middle) implying that the
positive (negative) E-field tends to weaken (strengthen) the
interaction between the hBN and the CNT. Because of the enhanced
electronic coupling between the CNT and hBN via the boron monovacancy
at the negative E-field, unwanted electrical conduction may occur
through the hBN sheet. We also observed that the charge distribution
near $E_F$ over the hBN is slightly modified and redistributed by the
E-field change. The charge redistibution affects the localized states
described above, and thus which are shifted up or down by the E-field
change.

For the $V_{\rm{N}}$ case, in contrast, it was found that the Fermi
level is aligned to the CBM of the CNT, which indicates that the CNT
becomes weakly electron-doped. Moreover, a very small localized peak
from the vacancy appears near $E_F$ as shown in Fig.~\ref{Fig3}(b).
This localized state is much less sensitive to the direction of the
applied E-field compared with the case of $V_{\rm{B}}$. In both
$V_{\rm{B}}$ and $V_{\rm{N}}$ cases, those localized states may result
in an electronic back scattering on the CNT surface, since they are
located near the VBM or the CBM of the CNT. 

For the case of $V_{\rm{B_3N}}$ in Fig.~\ref{Fig3}(c), the Fermi level
is located in the middle of the gap between the VBM and the CBM of the
CNT, and no charge transfer takes place between the CNT and the hBN
sheet. As shown in the PDOSs, all of the hBN states including defect
states originating from the vacancy are located far from $E_F$,
regardless of the E-field applied. Creating a $V_{\rm{B_3N}}$ vacancy
generates six N atoms surrounding the vacancy, each of which has
revealed a dangling bonds due to the reduction of its coordination
number from 3 to 2. However, these unsaturated bonds get all re-bonded
by the edge reconstruction resulting in no dangling bond. This is the
reason that no defect state occurs near $E_F$ of the system with
$V_{\rm{B_3N}}$. This implies that the presence of such
$V_{\rm{B_3N}}$ vacancies in hBN substrate may not influence the
electronic behaviors of CNT devices. In both $V_{\rm{B}}$ and
$V_{\rm{N}}$ cases, in contrast, there still remain unsaturated
dangling bonds at the N and B edge atoms enclosing the respective
vacancies corresponding to the defect states near $E_F$ mentioned
above.
 
More interestingly, magnetic properties give rise to the spin magnetic
moments calculated to be ${\mu}=1.57~{\mu_B}$ and ${\mu}=0$ for
$V_{\rm{B}}$ and $V_{\rm{N}}$, respectively. It was found that the
magnetic moments have been changed from those of their counterparts
without the CNT, which are both ${\mu}=1.00~{\mu_B}$~\cite{BingHuang}
meaning that there is apparently one unpaired electron at their
respective vacant sites. The results are in agreement with our Bader
charge analysis,~\cite{Bader} although it does not give an accurate
amount of charge transfer. ${\Delta}q$ is defined by the amount of
charge transfer to the hBN sheet from the CNT, and calculated to be
$0.49~e$ and $-0.73~e$ for $V_{\rm{B}}$ and $V_{\rm{N}}$,
respectively, where $e=-|e|(<0)$ is an electronic charge. Roughly
speaking, the charge of ${\sim}0.5~e$ transferred from the CNT
increases the spin magnetic moment for $V_{\rm{B}}$, and the electron
donation of ${\sim}1~e$ to the CNT removes the spin magnetic moment
for $V_{\rm{N}}$.

We also explored the dependence of magnetic moment on the applied
E-field. For $V_{\rm{B}}$, the magnetic moment increases (decreases)
by $0.18$~$\mu_B$ ($0.31$~$\mu_B$) to be ${\mu}=1.75~{\mu_B}$
(${\mu}=1.26~{\mu_B}$) from ${\mu}=1.57~{\mu_B}$ at an E-field value
of 0.2~V/{\AA} applied along the $+z$ ($-z$) direction. This E-field
dependence is also associated with ${\Delta}q$, which were calculated
to be ${\sim}0.2~e$ more (${\sim}0.3~e$ less) at an E-field value of
$+0.2$~V/{\AA} ($-0.2$~V/{\AA}) with respect to the case with no
E-field applied. For $V_{\rm{N}}$, which exhibits zero magnetic moment
without the applied E-field, we observed a revival of magnetic moment,
i.e., ${\mu}=0.80~{\mu_B}$, at an E-field value of $+0.2$~V/{\AA},
implying that 0.8 unpaired electrons have returned to the hBN. This is
consistent quite well with ${\Delta}q=-0.25~e$. At an E-field value of
$-0.2$~V/{\AA}, its magnetic moment is still zero keeping a complete
cancellation of one unpaired electron well matching our calculated
${\Delta}q$, which is closed to one. All values of the magnetic
moments are listed in Table~\ref{magmom}. Note that $V_{\rm{B_3N}}$
remains its nonmagnetic characteristic, regardless of the E-field
applied as well as of the presence of CNT.

\begin{figure}[t]
\includegraphics[width=1.0\columnwidth]{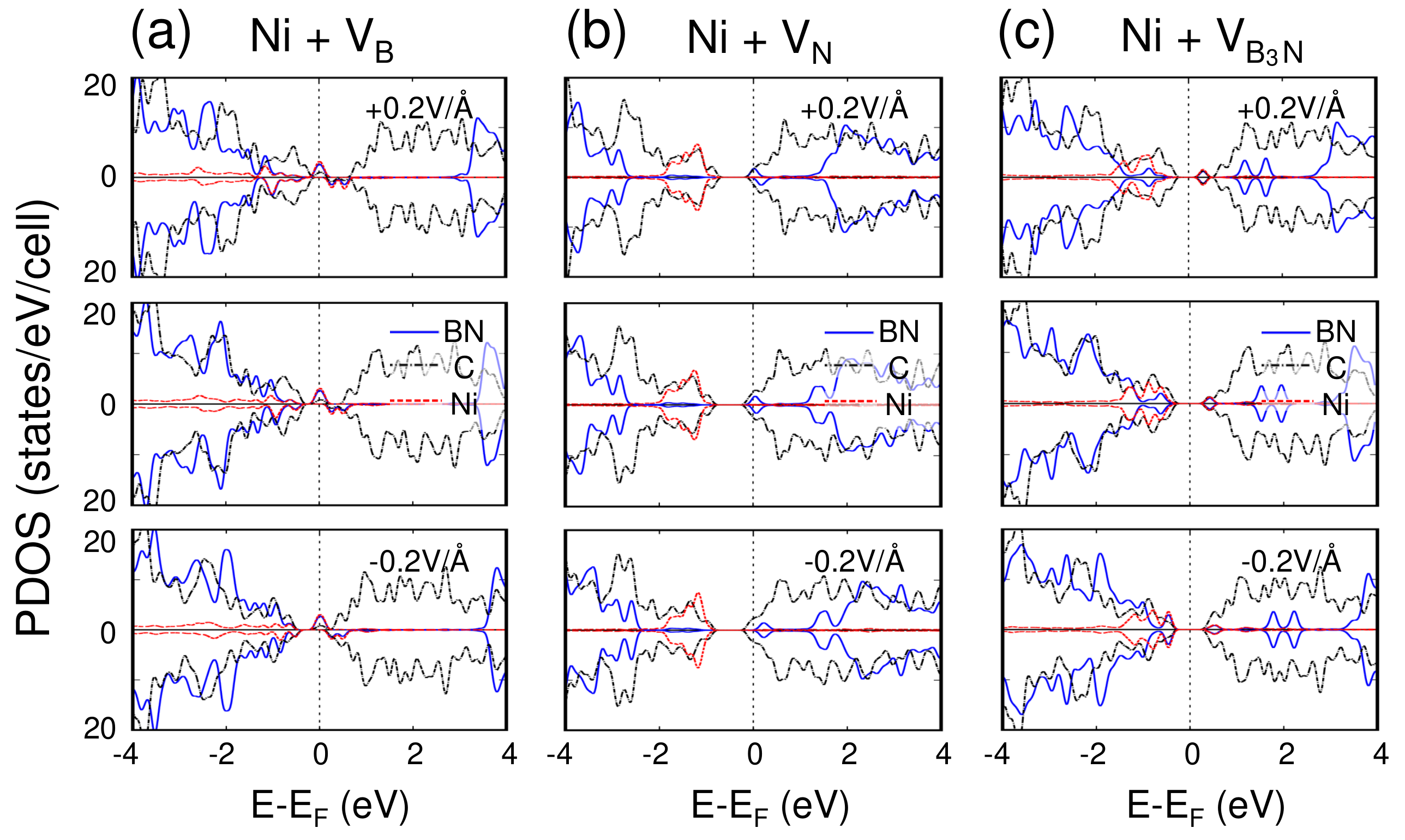}
\caption{(Color online) 
Spin-resolved PDOSs of the same configurations shown in
Fig.~\ref{Fig3}, but with a Ni impurity placed at (a) $V_{\rm{B}}$,
(b) $V_{\rm{N}}$, and (c) $V_{{\rm{B}}_3{\rm{N}}}$ in the absence
(middle) and the presence of an E-field applied with $+0.2$~V/{\AA}
(top) and $-0.2$~V/{\AA} along the $z$ direction, normal to the hBN
sheet. The Gaussian broadening of 0.15~eV was used for PDOSs.
\label{Fig4}}
\end{figure}

\begin{table}[bt]
\caption{Magnetic moments of undoped or Ni-doped system of a CNT on a
hBN sheet with a B or N monovacancy}
\label{magmom}
\begin{ruledtabular}
\begin{tabular}{r|ccc}
  \multirow{2}{*}{System} & \multicolumn{3}{c}{Magnetic moment ($\mu_B$)} \\
  & at $|{\bf{E}}|=+0.2$~V/{\AA} & $0.0$~V/{\AA} & $-0.2$~V/{\AA} \\
\hline
     $V_{\rm{B}}$ & 1.75 & 1.57 & 1.26 \\
     $V_{\rm{N}}$ & 0.80 & 0.00 & 0.00 \\
\hline
  Ni$+V_{\rm{B}}$ & 1.00 & 1.00 & 1.00 \\
  Ni$+V_{\rm{N}}$ & 0.54 & 0.11 & 0.00
\end{tabular}
\end{ruledtabular}
\end{table}


Finally, we added a Ni impurity to each of the three systems with
respective vacancies $V_{\rm{B}}$, $V_{\rm{N}}$, and $V_{\rm{B_3N}}$,
described above. Unlike in those systems without a Ni impurity
considered above, where the CNT is located directly above each vacant
site as shown in Fig.~\ref{Fig3}, the Ni adatom prefers to sit above
the center of each vacant site, and thus prevents the CNT from being
placed above the vacant site,  The equilibrium configuration of each
system is similar to that shown in Fig.~\ref{Fig1}(d).

Fig.~\ref{Fig4} shows the spin-resolved PDOSs of these three systems
obtained at three different E-field values, $+0.2$, $0.0$, and
$-0.2$~V/{\AA}. Their hBN states are shifted down (up) in energy under
the positive (negative) E-field, which is the same as seen in
Fig.~\ref{Fig2} and Fig.~\ref{Fig3}. The VBM and CBM of the hBN are
located far from $E_F$ for all the cases, but defect states are
produced by the Ni impurity as well as the vacancy. Especially for the
system with the Ni impurity on $V_{\rm{B}}$, it was observed in
Fig.~\ref{Fig4}(a) that some of these defect states are pinned at
$E_F$, and the states from the Ni atom (represented by a red dashed
line) are quite strongly hybridized with those from the boron
monovacancy (by a blue solid line) in a wide energy range in the
valence band. As mentioned earlier, the residual scattering in the
CNT-based device may be given rise to by the PDOS peak near $E_F$
formed by strong orbital hybridization between the Ni adatom, the CNT
and the hBN sheet. For the one with Ni$+V_{\rm{N}}$ or
Ni$+V_{\rm{B_3N}}$, on the other hand, its electronic structure and
its response to the E-field appear to be similar to the counterpart
system without a Ni impurity except for the states from the Ni
impurity mainly existing relatively deep inside the valence band. It
turns out in these two cases that the coupling strength between the Ni
impurity and the N or B$_3$N vacancy is relatively small, compared
with that in the Ni$+V_{\rm{B}}$ case.

We found that the Ni adatom takes part in determining the magnetic
properties of the systems. For $V_{\rm{B}}$, the Ni adatom reduces the
magnetic moment exactly to $1~\mu_B$ from $1.57~\mu_B$. This can be
also explained by our Bader charge analysis, which reveals that there
is almost no charge transfer to the CNT, but only between the hBN with
$V_{\rm{B}}$ and the Ni adatom. Those electrons participating in this
charge transfer remain as unpaired electrons at respective parts
keeping $\mu=1~\mu_B$. Moreover, its magnetic moment does not respond
to the E-field. Such an intriguing magnetic behavior is displayed in
the defect states pinned at $E_F$ shown in Fig.~\ref{Fig4}(a). On the
other hand, the Ni atdatom converts $V_{\rm{N}}$ to be magnetic with
the magnetic moment of $0.11~\mu_B$ at zero E-field, while maintaining
it nonmagnetic at $|{\bf{E}}|=-0.2$~V/{\AA}. According to our Bader
charge analysis, for Ni$+V_{\rm{N}}$, more electrons are donated to
the CNT than for $V_{\rm{N}}$, meaning a complete removal of unpaired
electrons from the hBN, but as an exception, at zero E-field, a small
amount of electrons is transferred to the hBN from the Ni adatom to
make the system weakly magnetic. We also observed that the Ni adatom
decreases the magnetic moment to $0.54~\mu_B$ from $0.80~\mu_B$ at an
E-field value of $+0.2$~V/{\AA}. The charge transfer from the hBN to
CNT is calculated to be $0.49~e$, which is ${\sim}0.24~e$ more than
that of the system without Ni adatom. All magnetic moment values are
summarized in Table~\ref{magmom}. $V_{\rm{B_3N}}$ remains nonmagnetic,
regardless of the existence of Ni impurity as well as E-field.

\section{Conclusions}
\label{Summary}

In summary, we have studied the effects of an external E-field and a
metal impurity on the electronic properties of CNTs on the hBN sheet
with and without vacancy defects. For each case, we obtained its
electronic structures such as the band structure, projected density of
states and local potential. We found that the electronic energy bands
of the hBN sheet are shifted in response to the applied E-field, 
regardless of whether the hBN is perfect or defective with a vacancy,
and whether there is a metal impurity or not. However, the shifted
electronic states in the valence and conduction bands of the perfect
hBN are located still far from $E_F$ under the field strength
considered, suggesting that the hBN sheet can be considered as a
suitable substrate material for CNT-based single-gate FETs, regardless
of the existence of a metal impurity and/or a B$_3$N vacancy. However,
hBN substrates with monovacancies and a metal impurity could exhibit
poor performance since the imperfections impair electrical
conductivity due to residual scattering when strong top-gate voltage
is applied in dual-gate FETs. Our theoretical results are in good
agreement with an experimental report that the in-gap states may
induce to facilitate residual scattering.~\cite{ETH}

\section*{Acknowledgments}

We acknowledge financial support from the Korean government through
National Research Foundation (NRF-2011-0016188 and 2013R1A1A2009131).
Some portion of our computational work was done using the resources of
the KISTI Supercomputing Center (KSC-2013-C1-031).

\bibliographystyle{apsrev}
\bibliography{BN-CNT} 

\begin{thebibliography}{46}
\expandafter\ifx\csname natexlab\endcsname\relax\def\natexlab#1{#1}\fi
\expandafter\ifx\csname bibnamefont\endcsname\relax
  \def\bibnamefont#1{#1}\fi
\expandafter\ifx\csname bibfnamefont\endcsname\relax
  \def\bibfnamefont#1{#1}\fi
\expandafter\ifx\csname citenamefont\endcsname\relax
  \def\citenamefont#1{#1}\fi
\expandafter\ifx\csname url\endcsname\relax
  \def\url#1{\texttt{#1}}\fi
\expandafter\ifx\csname urlprefix\endcsname\relax\def\urlprefix{URL }\fi
\providecommand{\bibinfo}[2]{#2}
\providecommand{\eprint}[2][]{\url{#2}}

\bibitem[{\citenamefont{Marchini et~al.}(2007)\citenamefont{Marchini,
  G\"unther, and Wintterlin}}]{GRARU}
\bibinfo{author}{\bibfnamefont{S.}~\bibnamefont{Marchini}},
  \bibinfo{author}{\bibfnamefont{S.}~\bibnamefont{G\"unther}},
  \bibnamefont{and}
  \bibinfo{author}{\bibfnamefont{J.}~\bibnamefont{Wintterlin}},
  \bibinfo{journal}{Phys. Rev. B} \textbf{\bibinfo{volume}{76}},
  \bibinfo{pages}{075429} (\bibinfo{year}{2007}).

\bibitem[{\citenamefont{N'Diaye et~al.}(2006)\citenamefont{N'Diaye, Bleikamp,
  Feibelman, and Michely}}]{GRAIR}
\bibinfo{author}{\bibfnamefont{A.~T.} \bibnamefont{N'Diaye}},
  \bibinfo{author}{\bibfnamefont{S.}~\bibnamefont{Bleikamp}},
  \bibinfo{author}{\bibfnamefont{P.~J.} \bibnamefont{Feibelman}},
  \bibnamefont{and} \bibinfo{author}{\bibfnamefont{T.}~\bibnamefont{Michely}},
  \bibinfo{journal}{Phys. Rev. Lett.} \textbf{\bibinfo{volume}{97}},
  \bibinfo{pages}{215501} (\bibinfo{year}{2006}).

\bibitem[{\citenamefont{Lui et~al.}(2009)\citenamefont{Lui, Liu, Mak, Flynn,
  and Heinz}}]{GRAMICA}
\bibinfo{author}{\bibfnamefont{C.~H.} \bibnamefont{Lui}},
  \bibinfo{author}{\bibfnamefont{L.}~\bibnamefont{Liu}},
  \bibinfo{author}{\bibfnamefont{K.~F.} \bibnamefont{Mak}},
  \bibinfo{author}{\bibfnamefont{G.~W.} \bibnamefont{Flynn}}, \bibnamefont{and}
  \bibinfo{author}{\bibfnamefont{T.~F.} \bibnamefont{Heinz}},
  \bibinfo{journal}{Nature} \textbf{\bibinfo{volume}{462}},
  \bibinfo{pages}{339} (\bibinfo{year}{2009}).

\bibitem[{\citenamefont{Berger et~al.}(2006)\citenamefont{Berger, Song, Li, Wu,
  Brown, Naud, Mayou, Li, Hass, Marchenkov et~al.}}]{SIC1}
\bibinfo{author}{\bibfnamefont{C.}~\bibnamefont{Berger}},
  \bibinfo{author}{\bibfnamefont{Z.}~\bibnamefont{Song}},
  \bibinfo{author}{\bibfnamefont{X.}~\bibnamefont{Li}},
  \bibinfo{author}{\bibfnamefont{X.}~\bibnamefont{Wu}},
  \bibinfo{author}{\bibfnamefont{N.}~\bibnamefont{Brown}},
  \bibinfo{author}{\bibfnamefont{C.}~\bibnamefont{Naud}},
  \bibinfo{author}{\bibfnamefont{D.}~\bibnamefont{Mayou}},
  \bibinfo{author}{\bibfnamefont{T.}~\bibnamefont{Li}},
  \bibinfo{author}{\bibfnamefont{J.}~\bibnamefont{Hass}},
  \bibinfo{author}{\bibfnamefont{A.~N.} \bibnamefont{Marchenkov}},
  \bibnamefont{et~al.}, \bibinfo{journal}{Science}
  \textbf{\bibinfo{volume}{312}}, \bibinfo{pages}{1191} (\bibinfo{year}{2006}).

\bibitem[{\citenamefont{Brar et~al.}(2007)\citenamefont{Brar, Zhang, Yayon,
  Ohta, McChesney, Bostwick, Rotenberg, Horn, and Crommie}}]{SIC2}
\bibinfo{author}{\bibfnamefont{V.~W.} \bibnamefont{Brar}},
  \bibinfo{author}{\bibfnamefont{Y.}~\bibnamefont{Zhang}},
  \bibinfo{author}{\bibfnamefont{Y.}~\bibnamefont{Yayon}},
  \bibinfo{author}{\bibfnamefont{T.}~\bibnamefont{Ohta}},
  \bibinfo{author}{\bibfnamefont{J.~L.} \bibnamefont{McChesney}},
  \bibinfo{author}{\bibfnamefont{A.}~\bibnamefont{Bostwick}},
  \bibinfo{author}{\bibfnamefont{E.}~\bibnamefont{Rotenberg}},
  \bibinfo{author}{\bibfnamefont{K.}~\bibnamefont{Horn}}, \bibnamefont{and}
  \bibinfo{author}{\bibfnamefont{M.~F.} \bibnamefont{Crommie}},
  \bibinfo{journal}{Appl. Phys. Lett.} \textbf{\bibinfo{volume}{91}},
  \bibinfo{pages}{122102} (\bibinfo{year}{2007}).

\bibitem[{\citenamefont{Novoselov et~al.}(2004)\citenamefont{Novoselov, Geim,
  Morozov, Jiang, Zhang, Dubonos, Grigorieva, and Firsov}}]{SIO21}
\bibinfo{author}{\bibfnamefont{K.}~\bibnamefont{Novoselov}},
  \bibinfo{author}{\bibfnamefont{A.}~\bibnamefont{Geim}},
  \bibinfo{author}{\bibfnamefont{S.}~\bibnamefont{Morozov}},
  \bibinfo{author}{\bibfnamefont{D.}~\bibnamefont{Jiang}},
  \bibinfo{author}{\bibfnamefont{Y.}~\bibnamefont{Zhang}},
  \bibinfo{author}{\bibfnamefont{S.}~\bibnamefont{Dubonos}},
  \bibinfo{author}{\bibfnamefont{I.}~\bibnamefont{Grigorieva}},
  \bibnamefont{and} \bibinfo{author}{\bibfnamefont{A.}~\bibnamefont{Firsov}},
  \bibinfo{journal}{Science} \textbf{\bibinfo{volume}{306}},
  \bibinfo{pages}{666} (\bibinfo{year}{2004}).

\bibitem[{\citenamefont{Zhang et~al.}(2005)\citenamefont{Zhang, Tan, Stormer,
  and Kim}}]{SIO22}
\bibinfo{author}{\bibfnamefont{Y.}~\bibnamefont{Zhang}},
  \bibinfo{author}{\bibfnamefont{Y.}~\bibnamefont{Tan}},
  \bibinfo{author}{\bibfnamefont{H.}~\bibnamefont{Stormer}}, \bibnamefont{and}
  \bibinfo{author}{\bibfnamefont{P.}~\bibnamefont{Kim}},
  \bibinfo{journal}{Nature} \textbf{\bibinfo{volume}{438}},
  \bibinfo{pages}{201} (\bibinfo{year}{2005}).

\bibitem[{\citenamefont{Zhang et~al.}(2008)\citenamefont{Zhang, Brar, Wang,
  Girit, Yayon, Panlasigui, Zettl, and Crommie}}]{SIO23}
\bibinfo{author}{\bibfnamefont{Y.}~\bibnamefont{Zhang}},
  \bibinfo{author}{\bibfnamefont{V.~W.} \bibnamefont{Brar}},
  \bibinfo{author}{\bibfnamefont{F.}~\bibnamefont{Wang}},
  \bibinfo{author}{\bibfnamefont{C.}~\bibnamefont{Girit}},
  \bibinfo{author}{\bibfnamefont{Y.}~\bibnamefont{Yayon}},
  \bibinfo{author}{\bibfnamefont{M.}~\bibnamefont{Panlasigui}},
  \bibinfo{author}{\bibfnamefont{A.}~\bibnamefont{Zettl}}, \bibnamefont{and}
  \bibinfo{author}{\bibfnamefont{M.~F.} \bibnamefont{Crommie}},
  \bibinfo{journal}{Nat. Phys.} \textbf{\bibinfo{volume}{4}},
  \bibinfo{pages}{627} (\bibinfo{year}{2008}).

\bibitem[{\citenamefont{Deshpande et~al.}(2009)\citenamefont{Deshpande, Bao,
  Miao, Lau, and LeRoy}}]{SIO24}
\bibinfo{author}{\bibfnamefont{A.}~\bibnamefont{Deshpande}},
  \bibinfo{author}{\bibfnamefont{W.}~\bibnamefont{Bao}},
  \bibinfo{author}{\bibfnamefont{F.}~\bibnamefont{Miao}},
  \bibinfo{author}{\bibfnamefont{C.~N.} \bibnamefont{Lau}}, \bibnamefont{and}
  \bibinfo{author}{\bibfnamefont{B.~J.} \bibnamefont{LeRoy}},
  \bibinfo{journal}{Phys. Rev. B} \textbf{\bibinfo{volume}{79}},
  \bibinfo{pages}{205411} (\bibinfo{year}{2009}).

\bibitem[{\citenamefont{Ishigami et~al.}(2007)\citenamefont{Ishigami, Chen,
  Cullen, Fuhrer, and Williams}}]{SIO25}
\bibinfo{author}{\bibfnamefont{M.}~\bibnamefont{Ishigami}},
  \bibinfo{author}{\bibfnamefont{J.~H.} \bibnamefont{Chen}},
  \bibinfo{author}{\bibfnamefont{W.~G.} \bibnamefont{Cullen}},
  \bibinfo{author}{\bibfnamefont{M.~S.} \bibnamefont{Fuhrer}},
  \bibnamefont{and} \bibinfo{author}{\bibfnamefont{E.~D.}
  \bibnamefont{Williams}}, \bibinfo{journal}{Nano Lett.}
  \textbf{\bibinfo{volume}{7}}, \bibinfo{pages}{1643} (\bibinfo{year}{2007}).

\bibitem[{\citenamefont{Martin et~al.}(2008)\citenamefont{Martin, Akerman,
  Ulbricht, Lohmann, Smet, Von~Klitzing, and Yacoby}}]{REDM}
\bibinfo{author}{\bibfnamefont{J.}~\bibnamefont{Martin}},
  \bibinfo{author}{\bibfnamefont{N.}~\bibnamefont{Akerman}},
  \bibinfo{author}{\bibfnamefont{G.}~\bibnamefont{Ulbricht}},
  \bibinfo{author}{\bibfnamefont{T.}~\bibnamefont{Lohmann}},
  \bibinfo{author}{\bibfnamefont{J.~H.} \bibnamefont{Smet}},
  \bibinfo{author}{\bibfnamefont{K.}~\bibnamefont{Von~Klitzing}},
  \bibnamefont{and} \bibinfo{author}{\bibfnamefont{A.}~\bibnamefont{Yacoby}},
  \bibinfo{journal}{Nat. Phys} \textbf{\bibinfo{volume}{4}},
  \bibinfo{pages}{144} (\bibinfo{year}{2008}).

\bibitem[{\citenamefont{Zhang et~al.}(2009)\citenamefont{Zhang, Brar, Girit,
  Zettl, and Crommie}}]{REDM1}
\bibinfo{author}{\bibfnamefont{Y.}~\bibnamefont{Zhang}},
  \bibinfo{author}{\bibfnamefont{V.~W.} \bibnamefont{Brar}},
  \bibinfo{author}{\bibfnamefont{C.}~\bibnamefont{Girit}},
  \bibinfo{author}{\bibfnamefont{A.}~\bibnamefont{Zettl}}, \bibnamefont{and}
  \bibinfo{author}{\bibfnamefont{M.~F.} \bibnamefont{Crommie}},
  \bibinfo{journal}{Nat. Phys.} \textbf{\bibinfo{volume}{5}},
  \bibinfo{pages}{722} (\bibinfo{year}{2009}).

\bibitem[{\citenamefont{Gannett et~al.}(2011)\citenamefont{Gannett, Regan,
  Watanabe, Taniguchi, Crommie, and Zettl}}]{IMMO}
\bibinfo{author}{\bibfnamefont{W.}~\bibnamefont{Gannett}},
  \bibinfo{author}{\bibfnamefont{W.}~\bibnamefont{Regan}},
  \bibinfo{author}{\bibfnamefont{K.}~\bibnamefont{Watanabe}},
  \bibinfo{author}{\bibfnamefont{T.}~\bibnamefont{Taniguchi}},
  \bibinfo{author}{\bibfnamefont{M.~F.} \bibnamefont{Crommie}},
  \bibnamefont{and} \bibinfo{author}{\bibfnamefont{A.}~\bibnamefont{Zettl}},
  \bibinfo{journal}{Appl. Phys. Lett.} \textbf{\bibinfo{volume}{98}},
  \bibinfo{pages}{242105} (\bibinfo{year}{2011}).

\bibitem[{\citenamefont{Dean et~al.}(2010)\citenamefont{Dean, Young, Meric,
  Lee, Wang, Sorgenfrei, Watanabe, Taniguchi, Kim, Shepard et~al.}}]{HQGRA}
\bibinfo{author}{\bibfnamefont{C.~R.} \bibnamefont{Dean}},
  \bibinfo{author}{\bibfnamefont{A.~F.} \bibnamefont{Young}},
  \bibinfo{author}{\bibfnamefont{I.}~\bibnamefont{Meric}},
  \bibinfo{author}{\bibfnamefont{C.}~\bibnamefont{Lee}},
  \bibinfo{author}{\bibfnamefont{L.}~\bibnamefont{Wang}},
  \bibinfo{author}{\bibfnamefont{S.}~\bibnamefont{Sorgenfrei}},
  \bibinfo{author}{\bibfnamefont{K.}~\bibnamefont{Watanabe}},
  \bibinfo{author}{\bibfnamefont{T.}~\bibnamefont{Taniguchi}},
  \bibinfo{author}{\bibfnamefont{P.}~\bibnamefont{Kim}},
  \bibinfo{author}{\bibfnamefont{K.~L.} \bibnamefont{Shepard}},
  \bibnamefont{et~al.}, \bibinfo{journal}{Nat. Nanotech.}
  \textbf{\bibinfo{volume}{5}}, \bibinfo{pages}{722} (\bibinfo{year}{2010}).

\bibitem[{\citenamefont{Jain et~al.}(2012)\citenamefont{Jain, Bansal, Durcan,
  and Yu}}]{exper1}
\bibinfo{author}{\bibfnamefont{N.}~\bibnamefont{Jain}},
  \bibinfo{author}{\bibfnamefont{T.}~\bibnamefont{Bansal}},
  \bibinfo{author}{\bibfnamefont{C.}~\bibnamefont{Durcan}}, \bibnamefont{and}
  \bibinfo{author}{\bibfnamefont{B.}~\bibnamefont{Yu}}, \bibinfo{journal}{IEEE
  Electron Device Lett.} \textbf{\bibinfo{volume}{33}}, \bibinfo{pages}{925}
  (\bibinfo{year}{2012}).

\bibitem[{\citenamefont{Baumgartner et~al.}(2014)\citenamefont{Baumgartner,
  Abulizi, Watanabe, Taniguchi, Gramich, and Schönenberger}}]{APL2014}
\bibinfo{author}{\bibfnamefont{A.}~\bibnamefont{Baumgartner}},
  \bibinfo{author}{\bibfnamefont{G.}~\bibnamefont{Abulizi}},
  \bibinfo{author}{\bibfnamefont{K.}~\bibnamefont{Watanabe}},
  \bibinfo{author}{\bibfnamefont{T.}~\bibnamefont{Taniguchi}},
  \bibinfo{author}{\bibfnamefont{J.}~\bibnamefont{Gramich}}, \bibnamefont{and}
  \bibinfo{author}{\bibfnamefont{C.}~\bibnamefont{Schönenberger}},
  \bibinfo{journal}{Appl. Phys. Lett.} \textbf{\bibinfo{volume}{105}}
  (\bibinfo{year}{2014}).

\bibitem[{\citenamefont{Xue et~al.}(2011)\citenamefont{Xue, Sanchez-Yamagishi,
  Bulmash, Jacquod, Deshpande, Watanabe, Taniguchi, Jarillo-Herrero, and
  Leroy}}]{STM1}
\bibinfo{author}{\bibfnamefont{J.}~\bibnamefont{Xue}},
  \bibinfo{author}{\bibfnamefont{J.}~\bibnamefont{Sanchez-Yamagishi}},
  \bibinfo{author}{\bibfnamefont{D.}~\bibnamefont{Bulmash}},
  \bibinfo{author}{\bibfnamefont{P.}~\bibnamefont{Jacquod}},
  \bibinfo{author}{\bibfnamefont{A.}~\bibnamefont{Deshpande}},
  \bibinfo{author}{\bibfnamefont{K.}~\bibnamefont{Watanabe}},
  \bibinfo{author}{\bibfnamefont{T.}~\bibnamefont{Taniguchi}},
  \bibinfo{author}{\bibfnamefont{P.}~\bibnamefont{Jarillo-Herrero}},
  \bibnamefont{and} \bibinfo{author}{\bibfnamefont{B.~J.} \bibnamefont{Leroy}},
  \bibinfo{journal}{Nat. Mater.} \textbf{\bibinfo{volume}{10}},
  \bibinfo{pages}{282} (\bibinfo{year}{2011}).

\bibitem[{\citenamefont{Decker et~al.}(2011)\citenamefont{Decker, Wang, Brar,
  Regan, Tsai, Wu, Gannett, Zettl, and Crommie}}]{STM2}
\bibinfo{author}{\bibfnamefont{R.}~\bibnamefont{Decker}},
  \bibinfo{author}{\bibfnamefont{Y.}~\bibnamefont{Wang}},
  \bibinfo{author}{\bibfnamefont{V.~W.} \bibnamefont{Brar}},
  \bibinfo{author}{\bibfnamefont{W.}~\bibnamefont{Regan}},
  \bibinfo{author}{\bibfnamefont{H.-Z.} \bibnamefont{Tsai}},
  \bibinfo{author}{\bibfnamefont{Q.}~\bibnamefont{Wu}},
  \bibinfo{author}{\bibfnamefont{W.}~\bibnamefont{Gannett}},
  \bibinfo{author}{\bibfnamefont{A.}~\bibnamefont{Zettl}}, \bibnamefont{and}
  \bibinfo{author}{\bibfnamefont{M.~F.} \bibnamefont{Crommie}},
  \bibinfo{journal}{Nano Lett.} \textbf{\bibinfo{volume}{11}},
  \bibinfo{pages}{2291} (\bibinfo{year}{2011}).

\bibitem[{\citenamefont{Watanabe et~al.}(2004)\citenamefont{Watanabe,
  Taniguchi, and Kanda}}]{BG}
\bibinfo{author}{\bibfnamefont{K.}~\bibnamefont{Watanabe}},
  \bibinfo{author}{\bibfnamefont{T.}~\bibnamefont{Taniguchi}},
  \bibnamefont{and} \bibinfo{author}{\bibfnamefont{H.}~\bibnamefont{Kanda}},
  \bibinfo{journal}{Nat. Mater.} \textbf{\bibinfo{volume}{3}},
  \bibinfo{pages}{404} (\bibinfo{year}{2004}).

\bibitem[{\citenamefont{Pierson}(1975)}]{BF3NH3CVD}
\bibinfo{author}{\bibfnamefont{H.~O.} \bibnamefont{Pierson}},
  \bibinfo{journal}{J. cpmpos. Mater.} \textbf{\bibinfo{volume}{9}},
  \bibinfo{pages}{228} (\bibinfo{year}{1975}).

\bibitem[{\citenamefont{Rozenberg et~al.}(1993)\citenamefont{Rozenberg,
  Sinenko, and Chukanov}}]{BClNH3CVD}
\bibinfo{author}{\bibfnamefont{A.~S.} \bibnamefont{Rozenberg}},
  \bibinfo{author}{\bibfnamefont{Y.~A.} \bibnamefont{Sinenko}},
  \bibnamefont{and} \bibinfo{author}{\bibfnamefont{N.~V.}
  \bibnamefont{Chukanov}}, \bibinfo{journal}{J. Mater. Sci.}
  \textbf{\bibinfo{volume}{28}}, \bibinfo{pages}{5528} (\bibinfo{year}{1993}).

\bibitem[{\citenamefont{Middleman}(1993)}]{B2H6NH3CVD}
\bibinfo{author}{\bibfnamefont{S.}~\bibnamefont{Middleman}},
  \bibinfo{journal}{Mater. Sci. Eng., A} \textbf{\bibinfo{volume}{163}},
  \bibinfo{pages}{135} (\bibinfo{year}{1993}).

\bibitem[{\citenamefont{Adams}(1981)}]{B3N3H6CVD}
\bibinfo{author}{\bibfnamefont{A.~C.} \bibnamefont{Adams}},
  \bibinfo{journal}{J. Electochem. Soc.} \textbf{\bibinfo{volume}{128}},
  \bibinfo{pages}{1378} (\bibinfo{year}{1981}).

\bibitem[{\citenamefont{Auwärter et~al.}(2004)\citenamefont{Auwärter, Suter,
  Sachdev, and Greber}}]{B3N3H3Cl3CVD1}
\bibinfo{author}{\bibfnamefont{W.}~\bibnamefont{Auwärter}},
  \bibinfo{author}{\bibfnamefont{H.~U.} \bibnamefont{Suter}},
  \bibinfo{author}{\bibfnamefont{H.}~\bibnamefont{Sachdev}}, \bibnamefont{and}
  \bibinfo{author}{\bibfnamefont{T.}~\bibnamefont{Greber}},
  \bibinfo{journal}{Chem. Mater.} \textbf{\bibinfo{volume}{16}},
  \bibinfo{pages}{343} (\bibinfo{year}{2004}).

\bibitem[{\citenamefont{Müller et~al.}(2005)\citenamefont{Müller, Stöwe, and
  Sachdev}}]{B3N3H3Cl3CVD2}
\bibinfo{author}{\bibfnamefont{F.}~\bibnamefont{Müller}},
  \bibinfo{author}{\bibfnamefont{K.}~\bibnamefont{Stöwe}}, \bibnamefont{and}
  \bibinfo{author}{\bibfnamefont{H.}~\bibnamefont{Sachdev}},
  \bibinfo{journal}{Chem. Mater.} \textbf{\bibinfo{volume}{17}},
  \bibinfo{pages}{3464} (\bibinfo{year}{2005}).

\bibitem[{\citenamefont{Constant and Feurer}(1981)}]{B3N3Cl3CVD}
\bibinfo{author}{\bibfnamefont{G.}~\bibnamefont{Constant}} \bibnamefont{and}
  \bibinfo{author}{\bibfnamefont{R.}~\bibnamefont{Feurer}},
  \bibinfo{journal}{J. Less-Common Met.} \textbf{\bibinfo{volume}{82}},
  \bibinfo{pages}{113} (\bibinfo{year}{1981}).

\bibitem[{\citenamefont{Shi et~al.}(2010)\citenamefont{Shi, Hamsen, Jia, Kim,
  Reina, Hofmann, Hsu, Zhang, Li, Juang et~al.}}]{FLBNCVD}
\bibinfo{author}{\bibfnamefont{Y.}~\bibnamefont{Shi}},
  \bibinfo{author}{\bibfnamefont{C.}~\bibnamefont{Hamsen}},
  \bibinfo{author}{\bibfnamefont{X.}~\bibnamefont{Jia}},
  \bibinfo{author}{\bibfnamefont{K.~K.} \bibnamefont{Kim}},
  \bibinfo{author}{\bibfnamefont{A.}~\bibnamefont{Reina}},
  \bibinfo{author}{\bibfnamefont{M.}~\bibnamefont{Hofmann}},
  \bibinfo{author}{\bibfnamefont{A.~L.} \bibnamefont{Hsu}},
  \bibinfo{author}{\bibfnamefont{K.}~\bibnamefont{Zhang}},
  \bibinfo{author}{\bibfnamefont{H.}~\bibnamefont{Li}},
  \bibinfo{author}{\bibfnamefont{Z.~Y.} \bibnamefont{Juang}},
  \bibnamefont{et~al.}, \bibinfo{journal}{Nano Lett.}
  \textbf{\bibinfo{volume}{10}}, \bibinfo{pages}{4134} (\bibinfo{year}{2010}).

\bibitem[{\citenamefont{Kim et~al.}(2012)\citenamefont{Kim, Hsu, Jia, Kim, Shi,
  Hofmann, Nezich, Rodriguez-Nieva, Dresselhaus, Palacios et~al.}}]{MonoBNCVD}
\bibinfo{author}{\bibfnamefont{K.~K.} \bibnamefont{Kim}},
  \bibinfo{author}{\bibfnamefont{A.}~\bibnamefont{Hsu}},
  \bibinfo{author}{\bibfnamefont{X.}~\bibnamefont{Jia}},
  \bibinfo{author}{\bibfnamefont{S.~M.} \bibnamefont{Kim}},
  \bibinfo{author}{\bibfnamefont{Y.}~\bibnamefont{Shi}},
  \bibinfo{author}{\bibfnamefont{M.}~\bibnamefont{Hofmann}},
  \bibinfo{author}{\bibfnamefont{D.}~\bibnamefont{Nezich}},
  \bibinfo{author}{\bibfnamefont{F.}~\bibnamefont{Rodriguez-Nieva},
  \bibfnamefont{Joaquin}},
  \bibinfo{author}{\bibfnamefont{M.}~\bibnamefont{Dresselhaus}},
  \bibinfo{author}{\bibfnamefont{T.}~\bibnamefont{Palacios}},
  \bibnamefont{et~al.}, \bibinfo{journal}{Nano Lett.}
  \textbf{\bibinfo{volume}{12}}, \bibinfo{pages}{161} (\bibinfo{year}{2012}).

\bibitem[{\citenamefont{Wang et~al.}(2013)\citenamefont{Wang, Jang, Jang, Kim,
  Park, Kim, Kahng, Choi, Ruoff, Song et~al.}}]{MWang}
\bibinfo{author}{\bibfnamefont{M.}~\bibnamefont{Wang}},
  \bibinfo{author}{\bibfnamefont{S.~K.} \bibnamefont{Jang}},
  \bibinfo{author}{\bibfnamefont{W.-J.} \bibnamefont{Jang}},
  \bibinfo{author}{\bibfnamefont{M.}~\bibnamefont{Kim}},
  \bibinfo{author}{\bibfnamefont{S.-Y.} \bibnamefont{Park}},
  \bibinfo{author}{\bibfnamefont{S.-W.} \bibnamefont{Kim}},
  \bibinfo{author}{\bibfnamefont{S.-J.} \bibnamefont{Kahng}},
  \bibinfo{author}{\bibfnamefont{J.-Y.} \bibnamefont{Choi}},
  \bibinfo{author}{\bibfnamefont{R.~S.} \bibnamefont{Ruoff}},
  \bibinfo{author}{\bibfnamefont{Y.~J.} \bibnamefont{Song}},
  \bibnamefont{et~al.}, \bibinfo{journal}{Adv. Mater}
  \textbf{\bibinfo{volume}{25}}, \bibinfo{pages}{2746} (\bibinfo{year}{2013}).

\bibitem[{\citenamefont{Jin et~al.}(2009)\citenamefont{Jin, Lin, Suenaga, and
  Iijima}}]{CJin}
\bibinfo{author}{\bibfnamefont{C.}~\bibnamefont{Jin}},
  \bibinfo{author}{\bibfnamefont{F.}~\bibnamefont{Lin}},
  \bibinfo{author}{\bibfnamefont{K.}~\bibnamefont{Suenaga}}, \bibnamefont{and}
  \bibinfo{author}{\bibfnamefont{S.}~\bibnamefont{Iijima}},
  \bibinfo{journal}{Phys. Rev. Lett.} \textbf{\bibinfo{volume}{102}},
  \bibinfo{pages}{195505} (\bibinfo{year}{2009}).

\bibitem[{\citenamefont{Song et~al.}(2010)\citenamefont{Song, Ci, Lu, Sorokin,
  Jin, Ni, Kvashnin, Kvashnin, Lou, Yakobson et~al.}}]{VACCVD}
\bibinfo{author}{\bibfnamefont{L.}~\bibnamefont{Song}},
  \bibinfo{author}{\bibfnamefont{L.}~\bibnamefont{Ci}},
  \bibinfo{author}{\bibfnamefont{H.}~\bibnamefont{Lu}},
  \bibinfo{author}{\bibfnamefont{P.~B.} \bibnamefont{Sorokin}},
  \bibinfo{author}{\bibfnamefont{C.}~\bibnamefont{Jin}},
  \bibinfo{author}{\bibfnamefont{J.}~\bibnamefont{Ni}},
  \bibinfo{author}{\bibfnamefont{A.~G.} \bibnamefont{Kvashnin}},
  \bibinfo{author}{\bibfnamefont{D.~G.} \bibnamefont{Kvashnin}},
  \bibinfo{author}{\bibfnamefont{J.}~\bibnamefont{Lou}},
  \bibinfo{author}{\bibfnamefont{B.~I.} \bibnamefont{Yakobson}},
  \bibnamefont{et~al.}, \bibinfo{journal}{Nano Lett.}
  \textbf{\bibinfo{volume}{10}}, \bibinfo{pages}{3209} (\bibinfo{year}{2010}).

\bibitem[{\citenamefont{Pumera}(2007)}]{IM1}
\bibinfo{author}{\bibfnamefont{M.}~\bibnamefont{Pumera}},
  \bibinfo{journal}{Langmuir} \textbf{\bibinfo{volume}{23}},
  \bibinfo{pages}{6453} (\bibinfo{year}{2007}).

\bibitem[{\citenamefont{Harris}(2007)}]{IM2}
\bibinfo{author}{\bibfnamefont{P.~J.} \bibnamefont{Harris}},
  \bibinfo{journal}{Carbon} \textbf{\bibinfo{volume}{45}}, \bibinfo{pages}{229
  } (\bibinfo{year}{2007}).

\bibitem[{\citenamefont{Takagi et~al.}(2006)\citenamefont{Takagi, Homma,
  Hibino, Suzuki, and Kobayashi}}]{IM3}
\bibinfo{author}{\bibfnamefont{D.}~\bibnamefont{Takagi}},
  \bibinfo{author}{\bibfnamefont{Y.}~\bibnamefont{Homma}},
  \bibinfo{author}{\bibfnamefont{H.}~\bibnamefont{Hibino}},
  \bibinfo{author}{\bibfnamefont{S.}~\bibnamefont{Suzuki}}, \bibnamefont{and}
  \bibinfo{author}{\bibfnamefont{Y.}~\bibnamefont{Kobayashi}},
  \bibinfo{journal}{Nano Lett.} \textbf{\bibinfo{volume}{6}},
  \bibinfo{pages}{2642} (\bibinfo{year}{2006}).

\bibitem[{\citenamefont{Banks et~al.}(2006)\citenamefont{Banks, Crossley,
  Salter, Wilkins, and Compton}}]{IM4}
\bibinfo{author}{\bibfnamefont{C.~E.} \bibnamefont{Banks}},
  \bibinfo{author}{\bibfnamefont{A.}~\bibnamefont{Crossley}},
  \bibinfo{author}{\bibfnamefont{C.}~\bibnamefont{Salter}},
  \bibinfo{author}{\bibfnamefont{S.~J.} \bibnamefont{Wilkins}},
  \bibnamefont{and} \bibinfo{author}{\bibfnamefont{R.~G.}
  \bibnamefont{Compton}}, \bibinfo{journal}{Angew. Chem. Int. Edn}
  \textbf{\bibinfo{volume}{45}}, \bibinfo{pages}{2533} (\bibinfo{year}{2006}).

\bibitem[{\citenamefont{Hofmann et~al.}(2007)\citenamefont{Hofmann, Sharma,
  Ducati, Du, Mattevi, Cepek, Cantoro, Pisana, Parvez, Cervantes-Sodi
  et~al.}}]{IM5}
\bibinfo{author}{\bibfnamefont{S.}~\bibnamefont{Hofmann}},
  \bibinfo{author}{\bibfnamefont{R.}~\bibnamefont{Sharma}},
  \bibinfo{author}{\bibfnamefont{C.}~\bibnamefont{Ducati}},
  \bibinfo{author}{\bibfnamefont{G.}~\bibnamefont{Du}},
  \bibinfo{author}{\bibfnamefont{C.}~\bibnamefont{Mattevi}},
  \bibinfo{author}{\bibfnamefont{C.}~\bibnamefont{Cepek}},
  \bibinfo{author}{\bibfnamefont{M.}~\bibnamefont{Cantoro}},
  \bibinfo{author}{\bibfnamefont{S.}~\bibnamefont{Pisana}},
  \bibinfo{author}{\bibfnamefont{A.}~\bibnamefont{Parvez}},
  \bibinfo{author}{\bibfnamefont{F.}~\bibnamefont{Cervantes-Sodi}},
  \bibnamefont{et~al.}, \bibinfo{journal}{Nano Lett.}
  \textbf{\bibinfo{volume}{7}}, \bibinfo{pages}{602} (\bibinfo{year}{2007}).

\bibitem[{\citenamefont{Kresse and Furthm\"uller}(1996)}]{VASP}
\bibinfo{author}{\bibfnamefont{G.}~\bibnamefont{Kresse}} \bibnamefont{and}
  \bibinfo{author}{\bibfnamefont{J.}~\bibnamefont{Furthm\"uller}},
  \bibinfo{journal}{Phys. Rev. B} \textbf{\bibinfo{volume}{54}},
  \bibinfo{pages}{11169} (\bibinfo{year}{1996}).

\bibitem[{\citenamefont{Kresse and Joubert}(1999)}]{PAW}
\bibinfo{author}{\bibfnamefont{G.}~\bibnamefont{Kresse}} \bibnamefont{and}
  \bibinfo{author}{\bibfnamefont{D.}~\bibnamefont{Joubert}},
  \bibinfo{journal}{Phys. Rev. B} \textbf{\bibinfo{volume}{59}},
  \bibinfo{pages}{1758} (\bibinfo{year}{1999}).

\bibitem[{\citenamefont{Ceperley and Alder}(1980)}]{LDACA}
\bibinfo{author}{\bibfnamefont{D.~M.} \bibnamefont{Ceperley}} \bibnamefont{and}
  \bibinfo{author}{\bibfnamefont{B.~J.} \bibnamefont{Alder}},
  \bibinfo{journal}{Phys. Rev. Lett.} \textbf{\bibinfo{volume}{45}},
  \bibinfo{pages}{566} (\bibinfo{year}{1980}).

\bibitem[{\citenamefont{Paszkowicz et~al.}(2002)\citenamefont{Paszkowicz,
  Pelka, Knapp, Szyszko, and Podsiadlo}}]{Paszkowicz2002}
\bibinfo{author}{\bibfnamefont{W.}~\bibnamefont{Paszkowicz}},
  \bibinfo{author}{\bibfnamefont{J.}~\bibnamefont{Pelka}},
  \bibinfo{author}{\bibfnamefont{M.}~\bibnamefont{Knapp}},
  \bibinfo{author}{\bibfnamefont{T.}~\bibnamefont{Szyszko}}, \bibnamefont{and}
  \bibinfo{author}{\bibfnamefont{S.}~\bibnamefont{Podsiadlo}},
  \bibinfo{journal}{Appl. Phys. A} \textbf{\bibinfo{volume}{75}},
  \bibinfo{pages}{431} (\bibinfo{year}{2002}).

\bibitem[{\citenamefont{Giovannetti et~al.}(2007)\citenamefont{Giovannetti,
  Khomyakov, Brocks, Kelly, and van~den Brink}}]{Giovannetti2007}
\bibinfo{author}{\bibfnamefont{G.}~\bibnamefont{Giovannetti}},
  \bibinfo{author}{\bibfnamefont{P.}~\bibnamefont{Khomyakov}},
  \bibinfo{author}{\bibfnamefont{G.}~\bibnamefont{Brocks}},
  \bibinfo{author}{\bibfnamefont{P.}~\bibnamefont{Kelly}}, \bibnamefont{and}
  \bibinfo{author}{\bibfnamefont{J.}~\bibnamefont{van~den Brink}},
  \bibinfo{journal}{Phys. Rev. B} \textbf{\bibinfo{volume}{76}},
  \bibinfo{pages}{073103} (\bibinfo{year}{2007}).

\bibitem[{\citenamefont{Liu et~al.}(2007)\citenamefont{Liu, Gurel, Morris,
  Murray, Zhitkovich, Kane, and Hurt}}]{NiCNT}
\bibinfo{author}{\bibfnamefont{X.}~\bibnamefont{Liu}},
  \bibinfo{author}{\bibfnamefont{V.}~\bibnamefont{Gurel}},
  \bibinfo{author}{\bibfnamefont{D.}~\bibnamefont{Morris}},
  \bibinfo{author}{\bibfnamefont{D.}~\bibnamefont{Murray}},
  \bibinfo{author}{\bibfnamefont{A.}~\bibnamefont{Zhitkovich}},
  \bibinfo{author}{\bibfnamefont{A.}~\bibnamefont{Kane}}, \bibnamefont{and}
  \bibinfo{author}{\bibfnamefont{R.}~\bibnamefont{Hurt}},
  \bibinfo{journal}{Adv. Mater.} \textbf{\bibinfo{volume}{19}},
  \bibinfo{pages}{2790} (\bibinfo{year}{2007}).

\bibitem[{\citenamefont{Park et~al.}(2014)\citenamefont{Park, Park, and
  Kim}}]{Park2014a}
\bibinfo{author}{\bibfnamefont{S.}~\bibnamefont{Park}},
  \bibinfo{author}{\bibfnamefont{C.}~\bibnamefont{Park}}, \bibnamefont{and}
  \bibinfo{author}{\bibfnamefont{G.}~\bibnamefont{Kim}}, \bibinfo{journal}{J.
  Chem. Phys.} \textbf{\bibinfo{volume}{140}}, \bibinfo{pages}{134706}
  (\bibinfo{year}{2014}).

\bibitem[{\citenamefont{Huang and Lee}(2012)}]{BingHuang}
\bibinfo{author}{\bibfnamefont{B.}~\bibnamefont{Huang}} \bibnamefont{and}
  \bibinfo{author}{\bibfnamefont{H.}~\bibnamefont{Lee}},
  \bibinfo{journal}{Phys. Rev. B} \textbf{\bibinfo{volume}{86}},
  \bibinfo{pages}{245406} (\bibinfo{year}{2012}).

\bibitem[{\citenamefont{Tang et~al.}(2009)\citenamefont{Tang, Sanville, and
  Henkelman}}]{Bader}
\bibinfo{author}{\bibfnamefont{W.}~\bibnamefont{Tang}},
  \bibinfo{author}{\bibfnamefont{E.}~\bibnamefont{Sanville}}, \bibnamefont{and}
  \bibinfo{author}{\bibfnamefont{G.}~\bibnamefont{Henkelman}},
  \bibinfo{journal}{J. Phys. Condens. Matter} \textbf{\bibinfo{volume}{21}},
  \bibinfo{pages}{084204} (\bibinfo{year}{2009}).

\bibitem[{\citenamefont{Droescher et~al.}(2012)\citenamefont{Droescher,
  Barraud, Watanabe, Taniguchi, Ihn, and Ensslin}}]{ETH}
\bibinfo{author}{\bibfnamefont{S.}~\bibnamefont{Droescher}},
  \bibinfo{author}{\bibfnamefont{C.}~\bibnamefont{Barraud}},
  \bibinfo{author}{\bibfnamefont{K.}~\bibnamefont{Watanabe}},
  \bibinfo{author}{\bibfnamefont{T.}~\bibnamefont{Taniguchi}},
  \bibinfo{author}{\bibfnamefont{T.}~\bibnamefont{Ihn}}, \bibnamefont{and}
  \bibinfo{author}{\bibfnamefont{K.}~\bibnamefont{Ensslin}},
  \bibinfo{journal}{New J. Phys.} \textbf{\bibinfo{volume}{14}},
  \bibinfo{pages}{103007} (\bibinfo{year}{2012}).

\end{thebibliography}
\end{document}